\documentclass{article}

\usepackage{arxiv}

\usepackage[utf8]{inputenc} % allow utf-8 input
\usepackage[T1]{fontenc}    % use 8-bit T1 fonts
\usepackage{hyperref}       % hyperlinks
\usepackage{url}            % simple URL typesetting
\usepackage{booktabs}       % professional-quality tables
\usepackage{amsfonts}       % blackboard math symbols
\usepackage{nicefrac}       % compact symbols for 1/2, etc.
\usepackage{microtype}      % microtypography
\usepackage{lipsum}
\usepackage[square,comma,numbers,sort&compress]{natbib}

\usepackage{color}

\title{Common Conversational Community Prototype: Scholarly Conversational Assistant}

\author{
Krisztian Balog \\
University of Stavanger\\
Norway \\
\texttt{\small krisztian.balog@uis.no} \\
\And
Lucie Flekova \\
Technische Universit{\"a}t Darmstadt\\
Germany\\
\texttt{\small l.flekova@gmail.com} \\
\And
Matthias Hagen \\
Martin-Luther-Universit{\"a}t Halle-Wittenberg \\
Germany \\
\texttt{\small matthias.hagen@informatik.uni-halle.de} \\
\AND
Rosie Jones \\
Spotify \\
United States \\
\texttt{\small rjones@spotify.com } \\
\And
Martin Potthast \\
Leipzig University \\
Germany \\
\texttt{\small martin.potthast@uni-leipzig.de} \\
\And
Filip Radlinski \\
Google \\
UK \\
\texttt{\small filiprad@google.com} \\
\AND
Mark Sanderson \\
RMIT University \\
Australia \\
\texttt{\small mark.sanderson@rmit.edu.au} \\
\And
Svitlana Vakulenko \\
University of Amsterdam \\
The Netherlands \\
\texttt{\small s.vakulenko@uva.nl} \\
\And
Hamed Zamani \\
Microsoft\\
United States\\
\texttt{\small hazamani@microsoft.com} \\
}

\begin{document}
\maketitle

\begin{abstract}
This paper discusses the potential for creating academic resources (tools, data, and evaluation approaches) to support research in conversational search, by focusing on realistic information needs and conversational interactions.
Specifically, we propose to develop and operate a prototype conversational search system for scholarly activities.
This Scholarly Conversational Assistant would serve as a useful tool, a means to create datasets, and a platform for running evaluation challenges by groups across the community.
This article results from discussions of a working group at Dagstuhl Seminar 19461 on Conversational Search.
\end{abstract}

% keywords can be removed

\section{Introduction}  % Motivation

Conversational search is a newly emerging research area that aims to provide access to digitally stored information by means of a conversational user interface, that is, a dialogue-based interaction inspired and informed by human communication processes~\cite{DBLP:conf/cui/2019,phdthesis_J,phdthesis_S}.
The major goal of a conversational search system is to effectively retrieve relevant answers to a wide range of questions expressed in natural language, with rich user-system dialogue as a crucial component for understanding the question and refining the answers~\cite{Allan:2012:FCO:2215676.2215678}.
The respective dialogue comprises of a sequence of exchanges between one or more users and a conversational search system, which can enable multi-step task completion and recommendation~\cite{Culpepper:2018:RFI:3274784.3274788}.
Several theoretical frameworks that further specify various components and requirements for an effective conversational search system have recently been proposed~\cite{Radlinski:2017:TFC:3020165.3020183,strathprints64619,Trippas:2017:PIC:3020165.3022144,DBLP:conf/ecir/VakulenkoRCR19,trippas2020towards}.

It is commonly recognized that only few natural conversational search corpora exist. Rather, corpora are often created through imagined needs (often in task-oriented Wizard-of-Oz studies), are inspired by logs, or come from crawls of community fora. This leads to significant research effort being planned around existing biased data and metrics, rather than data and metrics being constructed to support the most impactful research. While there have been instances of the research community interaction enabling research, such as at ECIR 2019,\footnote{\url{http://ecir2019.org/sociopatterns/}} this is relatively rare. One of our key motivations is to produce a system and corpus that contains and supports real user needs. 

Simultaneously, our community has common unsatisfied needs that appear very well suited to conversational search. Some common tasks are performed by researchers repeatedly without providing any community research value in terms of data and feedback collection, despite being relevant to many published experiments. Examples of these tasks include PC selection or finding interest profiles in EasyChair, or identifying the most relevant sessions in the Whova conference app. The collective time spent (arguably inefficiently) by our community on such tasks may far surpass the cost of creating a system that also supports research progress while providing this \emph{community value}.

\section{Proposed Research}

We propose to develop and operate a prototype conversational search system (Scholarly Conversational Assistant) that would serve as 
\begin{itemize}
\item a useful search tool, 
\item a means to create datasets for further academic research, 
\item and a platform for running evaluation challenges by groups across the community.
\end{itemize}

In particular, the Scholarly Conversational Assistant would allow our research community to perform a range of research-related activities. In extensive discussions, we settled on this domain for a number of reasons: (1) The data that is involved (such as papers authored, conferences/talks attended, PC memberships) is generally considered less private. Indeed most such data is already public albeit difficult to search. (2) The system is one that the members of our community would be using ourselves, giving an active knowledgeable participant base, who could contribute improvements and publish papers based on interactions observed. (3) It caters to a broad range of information needs (see below) that are currently not supported well by existing systems. (4) The relevant research groups could avoid competing with commercial providers.

A number of other possible domains were discussed, including movies, music, news, and podcasts. They have a significantly larger potential audience, yet potentially compete with commercial providers. In determining our plan, it became clear that some participants also consider interests in these areas to be highly sensitive or personal. As a critical constraint, privacy of relevant data is key (having impacted, for example, the Living Labs research \cite{Hopfgartner:2019:CEL} despite significant effort).

\section{Research Challenges}

The aim of the Scholarly Conversational Assistant system would be to enable a wide variety of research in conversational search by covering example information needs like:
\begin{itemize}
  \item ``What should I read?'' --- Find research on a new area of interest.
  \item ``Help me plan my attendance'' --- Plan what sessions to attend and whom to talk to at a conference. (Conference organizers could also use that information for optimizing room allocations.)
  \item ``Whom should I invite?'' --- Find conference PC, SPC, session chairs, invite speakers, etc.
\end{itemize}
Importantly, the system would log all interactions such that classes of information needs that have potential for study may be identified over time. People may evaluate the system by filling out a questionnaire, with the option of free text feedback, after each conversation (and possibly leave comments behind for individual system utterances).

\subsection{Connection to Knowledge Graphs}

The system would operate on a \emph{personal research graph} (PKG)~\cite{Balog:2019:PKG}, more specifically, the portion of the PKG that the user wants to share with the system. The PKG could include, among other information:
\begin{itemize}
  \item Authorship information (which may be connected to a public citation graph),
  \item Conference committee membership, awards, etc.,
  \item Talks given anywhere public,
  \item Attendance of conferences, sessions, etc.,
  \item (in the private part) Annotations of papers, notes on talks, etc.
\end{itemize}

\subsection{First Steps}

The project is ambitious, but we think it can be grown incrementally:
\begin{itemize}
  \item A starting point would be to get one ore more graduate students to start coding a tool and check it in to GitHub. It is likely that students will be able to build on top of existing infrastructure. In order for this to work, it will be necessary for a research team to own the decisions who (believes they will) get value out of such work. With a prototype system in place, one could establish a shared task at a workshop or conduct a lab study at scale. One might also design a challenge at TREC/CLEF to make use of the skeleton.
  \item One might alternatively start by collecting evidence that such a system is something the community actually wants. Here, a sample of dialogues or information needs (that one might want to support) could be gathered.
\end{itemize}

\section{Broader Impact}

The organization of shared tasks has a long tradition in information retrieval as well as natural language processing and the dialogue community within it. In conversational search, these two communities will collaborate to build search systems that have a natural language interface as well as conversational capabilities. The breadth of potential tasks that are due to this confluence of research fields---as also identified in Dagstuhl seminar 19461---is large. 
% FR: I don't think that "risk of fracture" is something we should be addressing. 
%     Researchers can collaborate or not as they wish to.
% REMOVED: This diversity poses a distinct risk to fracture the community, since a common understanding of how to evaluate conversational search systems and its components may not emerge quickly, unless there is some structuring influence. Shared tasks are a tried and tested solution in this respect.
% REPLACED WITH:
As such, developing common infrastructure and shared tasks would have high value for the community.

In particular, the outcome of shared tasks are typically large corpora and performance measures that, together, form reusable benchmarks. For example, the Cranfield-style evaluation frameworks that were adapted by TREC, or the corpora developed for the CoNLL shared tasks have had a broad impact on their respective communities at large. We expect that a conversational search challenge, too, will help to align and shape the community. 

Moreover, by developing specific shared tasks in the form of living labs~\cite{Hopfgartner:2018:ECS,Hopfgartner:2019:CEL}, we see the opportunity to apply early conversational search systems in practice as soon as possible. Here, the application domain of scholarly search, while allowing for a wide range of basic and advanced evaluation setups, may ideally transfer directly into new prototypes to enhance research itself, for instance, impacting the productivity of managing one’s personal conferences schedules.

\section{Obstacles and Risks}

A variety of systems for storing and accessing research publications, reviews and conference attendance already exist. For the Scholarly Conversational Assistant to be successful, it must either be more useful than these, or potentially integrate with them. Some of the existing systems include: dblp, semantic scholar, ACM library, Google scholar, ACL anthology, open review, arXiv, Athena conference chatbot, Citeseer, Arnetminer, and arXivDigest (more on these in related reading). 

Risks involved in operationalizing our envisaged conversational search system include:
\begin{itemize}
  \item \emph{Privacy and data retention rules.} Ideally, the Scholarly Conversational Assistant would allow the logging of user interactions including voice input. For all personal data, the system would require a process for data access, retention and deletion as well as logging, in compliance with local regulations. Even the use of third-party speech recognizers may be sensitive depending on the location of data storage. 
  \item \emph{Opinions != facts in indexing.}  Some information that could be collected is likely to be expressed opinions rather than facts (e.g., tweets about papers). Thus, we may want to allow verification of such information before use for search and recommendation, or present it in a separate clearly-marked format with the potential for correction or deletion. Others may wish to combine private information (such as a user’s personal opinions about papers), without this information being propagated. 
  \item \emph{Speech recognition.} The use of third-party speech recognizers may be sensitive depending on the location of data storage. In addition, in the Scholarly Conversational Assistant case, the corpus contains many proper names and technical terms. A speech recognizer may require a custom language model integrating this corpus to perform well.
  \item \emph{Personal Knowledge Graph implementation.} We would need a design that allows both cloud- and client-side storage of personal data. We need to make sure that private parts of the PKG remain private and also that users have full control over what is stored in their PKG. In case an offline dataset is created and shared, there needs to be an agreement in place that ensures that personal data would need to be removed upon request. (It should be noted that there is no way to enforce this, and ``unauthorized'' access may only be spotted if people publish using that data.)
  \item \emph{Usage volume.} Low user participation is a concern. Beyond ensuring that the system is useful, other ways to mitigate this could include rewarding (paying) users or incentivizing them through gamification (e.g., at conferences to use the system).
  \item \emph{Implementation.} The underlying system would require a significant effort to implement. As this would likely be contributions from different practitioners at various stages in their careers over an extended time, the contributors would naturally change. To alleviate some associated risk, a strong modularization would be beneficial, with clear interfaces and documentation. Moreover, the design of the initial prototype should be as simple as possible, with agreement of how the system’s continued development is ensured during operation. The live service would also need coordination, for example, of how live experiments are planned and executed.
  \item \emph{Operation.} Past academic systems have often been deployed on individual servers without redundancy, and potentially lacking resources for scalability. This project would likely wish to consider for this project to identify possible sponsorship from a cloud provider or host institution with significant cluster resources. The hosting decision should likely take into account long-term commitment.
  % REMOVED BY FR: For long-term support, the ideal solution would be to set up a government-funded not-for-profit organization or project to support continued operation.
  \item \emph{Stability and reproducibility.} If used for online challenges where participants submit code that runs live, this would need to be of suitable quality to be widely used. Care would need to be taken in designing common APIs that minimize the risks involved where a component does not behave as expected.
  % REWORDED BY FR: to the challenge that are prototypical in nature, and that are not necessarily fit for production use. In this respect, a slick interface needs to be defined that gives participants the ability to solve a specific problem whereas alleviating the from all other housekeeping required. Moreover, the systems should be collected and hosted at the same infrastructure where the evaluation framework itself runs to ensure frictionless operation.
\end{itemize}

\section{Suggested Readings and Resources}

In the following, we list a set of resources (data and tools) that might be useful in building such a system. 
%
%\begin{itemize}
% \item 

Software platforms:
 \begin{itemize}
  \item Macaw: A conversational information seeking platform implemented in Python which supports multiple interfaces and modalities~\cite{Zamani:2019}. 
  \item TIRA Integrated Research Architecture \cite{Potthast:2019} (a  modularized  platform  for  shared  tasks).
  \end{itemize}
%  \item 
Scientific IR tools:
\begin{itemize}  
  \item ArXivDigest: A personalized scientific literature recommendation framework based on arXiv articles.\footnote{\url{https://github.com/iai-group/arxivdigest}}
   \item GrapAL: Querying Semantic Scholar's literature graph \cite{Betts:2019} (web-based tool for exploring scientific literature, e.g., finding experts on a given topic).\footnote{\url{https://allenai.github.io/grapal-website/}}
    \end{itemize}

Open-source scholarly conversational agents:
    \begin{itemize}
  \item UKP-ATHENA: A scientific conversational agent \cite{Mesgar:2019} (early prototype for assisting ACL{*} conference attendees and answering basic ACL Anthology queries).\footnote{\url{http://athena.ukp.informatik.tu-darmstadt.de:5002/}}
  \end{itemize}
%\end{itemize}
%
Data collections suitable to be incorporated in the Scholarly Conversational Assistant include, but are not limited to:
\begin{itemize}
  \item Open Research Knowledge Graph\footnote{\url{http://orkg.org}} (ORKG)~\cite{Jaradeh:2019:ORK:3360901.3364435}: Semantic annotations of scientific publications 
  \item Semantic Scholar: Articles in a broad range of fields
  \item ACM DL: A subset of computer science articles
  \item dblp: A clean list of computer science articles
  \item ACL Anthology: A public collection of ACL{*} articles
  \item Open Review: A small subset of conference articles with public reviews
  \item Other sources include: Google Scholar, Citeseer, Arnetminer, and Conference attendance apps (e.g., Whova)
\end{itemize}
Other related work:
\begin{itemize}
  \item \citet{Gentile:2015:CLA}: Recupero: Conference Live: Accessible and Sociable Conference Semantic Data
  \item \citet{Dalton:2018:VGC}: Vote Goat: Conversational Movie Recommendation
  \item \citet{wan2019aminer}: Aminer: Search and mining of academic social networks (researcher-centric IR)
\end{itemize}

\bibliographystyle{abbrvnat}  
\bibliography{references}

\end{document}